# Theory of Quantum Pulse Position Modulation and Related Numerical Problems

G. Cariolaro, *Life Member, IEEE* and G. Pierobon, *Life Member, IEEE*

***Abstract*— The paper deals with quantum pulse position modulation (PPM), both in the absence (pure states) and in the presence (mixed states) of thermal noise, using the Glauber representation of coherent laser radiation. The objective is to find optimal (or suboptimal) measurement operators and to evaluate the corresponding error probability. For PPM, the correct formulation of quantum states is given by the tensorial product of $m$ identical Hilbert spaces, where $m$ is the PPM order. The presence of mixed states, due to thermal noise, generates an optimization problem involving matrices of huge dimensions, which already for 4–PPM, are of the order of ten thousand. To overcome this computational complexity, the currently available methods of quantum detection, which are based on explicit results, convex linear programming and square root measurement, are compared to find the computationally less expensive one. In this paper a fundamental role is played by the geometrically uniform symmetry of the quantum PPM format. The evaluation of error probability confirms the vast superiority of the quantum detection over its classical counterpart.**

## I. Introduction

Coherent optical communications represent the best way to convey an enormous amount of information over large distances, both through fibers and in free space. Quantum optical communications offer the possibility to improve the performance of (classical) optical communications by basing the design of the detection on the laws of Quantum Mechanics. Several authors have shown, both theoretically [3][16][24] and experimentally [4][5][19][23], that improvements, of at least a decade, can be obtained in terms of error probability. The uncertainty in quantum detection is intrinsically linked to the laws of quantum mechanics. In the language of classical optical systems, this uncertainty corresponds to the so called *shot noise*. Another source of uncertainty is the presence of *thermal noise*.

Necessary and sufficient conditions for the optimal measurement set were found in pioneering papers by Holevo [6] and Yuen *et al.* [7]. Although the optimal measurement set is completely characterized, explicit solutions for the optimal measurement set are in general not available. It is therefore necessary to resort to a numerical evaluation based on convex semidefinite programming [20]. However, under a certain symmetry constraint, called *geometrically uniform symmetry* (GUS), a simple measurement, introduced by Hausladen *et al.* [12], and known as *square root measurement* (SRM), turns out to be optimum. This SRM has the remarkable advantage that, starting from the given states, it can be straightforwardly evaluated. Moreover, even when it is not optimal, the SRM often provides "pretty good" upper bounds on the performance of optimal detectors.

The problem of quantum detection in a noisy environment has received scarce attention in the literature, but the extension of SRM to mixed states [21] has opened new perspectives. In a previous paper [24], we applied this extension to the analysis of quantum communication systems, in the presence of thermal noise, using the Glauber theory on coherent states [1] to represent the noisy quantum channel.

In this paper, the theory of [24], which was applied to PSK (phase shift keying) and QAM (quadrature amplitude modulation) systems, will be extended to PPM (quantum pulse position modulation) systems. The novelty of this modulation format is that it must be formulated (see [7][19]) in a composite Hilbert space $\mathcal{H}$, given by the tensor product $\mathcal{H}_0^{\otimes m}$ of $m$ identical Hilbert spaces $\mathcal{H}_0$, where $m$ is the PPM's alphabet size. Thus, the quantum states corresponding to the classical binary PPM symbols are given by the tensor product of $m$ Glauber coherent states. For instance, in 4–PPM, the composite states representing the symbols $[0001], [0010], [0100], [1000]$, are given by

$$\begin{array}{ll} |\gamma_0\rangle = |0\rangle\otimes|0\rangle\otimes|0\rangle\otimes|\alpha\rangle & |\gamma_1\rangle = |0\rangle\otimes|0\rangle\otimes|\alpha\rangle|\otimes 0\rangle \\ |\gamma_2\rangle = |0\rangle\otimes|\alpha\rangle\otimes|0\rangle\otimes|0\rangle & |\gamma_3\rangle = |\alpha\rangle\otimes|0\rangle\otimes|0\rangle\otimes|0\rangle \end{array} \quad (1)$$

where the symbol 1 is represented by a Glauber state $|\alpha\rangle$, with an average number of photons given by $N_s = |\alpha|^2$, and the symbol 0 by the Glauber "ground state" $|0\rangle$, which has zero photons. In the presence of noise (mixed states) the $m$ composite states must be replaced by $m$ composite density operators. If $n$ is the dimension of the basic Hilbert space $\mathcal{H}_0$, the dimension of the composite space $\mathcal{H} = \mathcal{H}_0^{\otimes m}$ becomes $N = n^m$, which gives rise to a problem of computational complexity. In the approximation of the Glauber representation, in order to guarantee an adequate approximation, the choice of dimension $n$ (theoretically infinite) depends on the average number of photons $N_s$, which in turn depends on the range of the error probability $P_e$ that we wish to explore. To fix ideas, we anticipate the following schedule for 4–PPM

- $P_e \simeq 10^{-2} \;\to\; N_s \simeq 3 \;\to\; n \simeq 10\,,\; N = n^m \simeq\; 10^4$
- $P_e \simeq 10^{-3} \;\to\; N_s \simeq 4.5 \to\; n \simeq 15\,,\; N = n^m \simeq\; 5\;10^4$
- $P_e \simeq 10^{-4} \;\to\; N_s \simeq 6.5 \to\; n \simeq 20\,,\; N = n^m \simeq 16\;10^4$

Thus, we realize that, while the dimension $n$ of basic space can be confined to moderately small values, the dimension $N$ of composite space quickly becomes huge with consequently

The authors are with Department of Information Engineering, University of Padova, Via Gradenigo 6/B - 35131 Padova, Italy.

severe numerical problems, mainly related to the eigendecomposition (EID) of $N \times N$ matrices. Therefore, one of the main targets of the paper will be a systematic comparison of the currently available methods, to find the lowest computational complexity and, where possible, to search for analytical results for which the numerical evaluation does not represent a problem for any arbitrarily large dimension (see the EID of the PPM symmetry operator in Section III).

The above schedule refers to a moderate amount of thermal noise ($\mathcal{N} = 0.05$ thermal noise photons). Note that in the absence of noise, that is with pure states, the analysis is straightforward and the error probability is known, in closed form, from the pioneering works of Yuen, Kennedy and Lax [7], although the authors do not mention applications to PPM.

This paper is organized as follows. In Section II we give a detailed overview of the quantum detection problem. In Section III we formulate the theory of PPM and prove that the quantum PPM format satisfies the GUS property, which surprisingly enough, has not been noticed elsewhere; the main task is to find an analytical expression for the symmetry operator and its EID. In Section IV we formulate the PPM quantum states according to the Glauber theory. Finally, in Section V, after a comparative discussion of computational complexity, we evaluate the PPM performances, up to the order $m = 4$. In Section VI we mention to the problem of implementation.

## II. Overview of Quantum Detection

This overview is finalized to PPM. For a general overview, see [24].

### A. Formulation of the problem

The quantum detection problem is stated as follows [3]. A transmitter conveys classical information to a receiver through a quantum–mechanical channel, by choosing a quantum state from a set $\{\rho_i | i = 0, \dots, m-1\}$ of *density operators* on an $N$–dimensional Hilbert space $\mathcal{H}$. The density operators $\rho_i$ are chosen with given *prior probabilities* $q_i$, but in what follows, we assume equal probabilities, i.e., $q_i = 1/m$. The receiver detects the transmitted information with a set of *positive operator valued measurements* (POVM) $\Pi_0, \dots, \Pi_{m-1}$. Provided that the state density operator is $\rho_i$, the probability that the detection system reveals the state $j$, is given by the transition probability $p(j|i) = \mathrm{Tr}(\rho_i \Pi_j)$, $i, j = 0, \dots, m-1$. Then, the probability of correct detection becomes

$$P_c = \frac{1}{m} \sum_{i=0}^{m-1} p(i|i) = \frac{1}{m} \sum_{i=0}^{m-1} \mathrm{Tr}(\rho_i \Pi_i) \, . \qquad (2)$$

The description through density operators represents the general case, and includes the pure state case, in which $\rho_i$ reduces to the rank–one operator $\rho_i = |\gamma_i\rangle\langle\gamma_i|$.

Quantum detection is simplified and has specific properties if the state constellation exhibits the Geometrically Uniform Symmetry (GUS), that is if a unitary operator $S$ exists with the properties[1]

$$\rho_i = S^i \rho_0 S^{-i} \, , \qquad S^m = I_{\mathcal{H}} \qquad (3)$$

where $S$ is the *symmetry operator*, and $\rho_0$ is the *generating density operator*. The simplifications with GUS are: 1) The quantum source is completely specified by $\rho_0$ and $S$. 2) The POVM $\Pi_i$ can be chosen to have the same GUS as the density operators $\rho_i$ [21], namely, $\Pi_i = S^i \, \Pi_0 \, S^{-i}$. The search can then be limited to the reference POVM $\Pi_0$. 3) The transition probabilities $p(j|i)$ depend only on the difference $i - j$, mod $m$ (their matrix is circulant). In fact, $p(j|i) = \mathrm{Tr}\left[\Pi_0 \, S^{i-j} \, \rho_0 \, S^{-(i-j)}\right]$. Consequently

$$P_c = \mathrm{Tr}(\rho_0 \Pi_0) \, . \qquad (4)$$

The positive semidefinite (PSD) Hermitian operators $\rho_0$ and $\Pi_0$ may, in some respect, be redundant and can then be investigated through their *factors*. In an $N$–dimensional Hilbert space the density operators $\rho_0$ are represented by $N \times N$ complex matrices and can be factorized in the form $\rho_0 = \gamma_0 \gamma_0^*$, where $\gamma_0$ is a convenient $N \times r_0$ matrix, with $r_0 = \mathrm{rank}\,(\rho_0)$. This factorization is not unique, but the ambiguity turns out to be irrelevant in quantum detection [21]. In particular, if the states are pure, then $r_0 = 1$ and $\gamma_0$ reduces to the ket $|\gamma_0\rangle$. Since the rank of the optimal POVM $\Pi_0$ is not greater than $r_0$ [21], we postulate a similar factorization $\Pi_0 = \mu_0 \mu_0^*$ of the measurement operator. In the presence of GUS the factors of $\rho_i$ and $\Pi_i$ can be obtained from the reference factors as $\gamma_i = S^i \, \gamma_0$ and $\mu_i = S^i \, \mu_0$, where $\gamma_i$ and $\mu_i$ have the same dimension $N \times r_0$.

### B. Optimal solution (minimum error probability)

In their pioneering papers, Holevo [6] and Yuen *et al.* [7] found necessary and sufficient conditions (see also [21]) for an optimum POVM set (which minimized the error probability). In a few special cases, the optimal POVM set is explicitly known, namely: 1) for binary systems in the general case of mixed states, and 2) for $m$-ary systems with pure states having GUS. Case 2) will be seen below in the context of SRM. Case 1) is due to Helstrom [3], who obtained the following explicit result:

Proposition 1 *Let $\rho_0$ and $\rho_1$ be the density operators of a binary quantum system. Define the decision operator $D = q_1 \rho_1 - q_0 \rho_0$, where $q_0$ and $q_1$ are the prior probabilities, and find the corresponding EID: $D = \sum_i \epsilon_i |\epsilon_i\rangle\langle\epsilon_i|$, where the eigenvalues $\epsilon_i$ are real and $|\epsilon_i\rangle$ are the corresponding eigenvectors. Then, the optimal POVM are given by $\Pi_0 = \sum_{\epsilon_i \le 0} |\epsilon_i\rangle\langle\epsilon_i|$ and $\Pi_1 = \sum_{\epsilon_i > 0} |\epsilon_i\rangle\langle\epsilon_i|$. The corresponding correct detection probability is given by $P_c = q_0 + \sum_{\epsilon_i > 0} \epsilon_i$.*

In the general case, explicit solutions for the optimal POVM are not available. However, the problem of maximizing the correct detection probability $P_c$ can be solved numerically by *convex semidefinite programming* [20]. For instance, one can use the LMI (Linear Matrix Inequality) Toolbox of MatLab, which permits evaluating the optimal performance with any desired accuracy. Of course, if the dimensions are very large, as in the case of PPM with mixed states, the numerical evaluation may become very time and storage consuming.

[1] In this formulation equal prior probabilities are implied. In the general case, the *weighted* density operators $q_i \rho_i$ should be considered [21]. Note that if the GUS holds with equal probabilities, as in (3), it does not hold with arbitrary probabilities.

### C. SRM solution (square root measurement)

A more straightforward, although not optimal, approach to the problem is offered by the SRM. The approach for pure states was proposed by Hausladen *et al.* [12] and thoroughly investigated by Eldar and Forney [18]. The generalization to mixed states is due to Eldar *et al.* [21]. In the last paper it is shown that SRM is equivalent to LSM (least square measurement). The SRM formulation is stated in terms of the *state matrix* $\Gamma = [\,\gamma_0, \gamma_1, \dots, \gamma_{m-1}\,]$ and of the *measurement matrix* $M = [\,\mu_0, \mu_1, \dots, \mu_{m-1}\,]$, both of dimension $N \times m\, r_0$. Now, the quantum detection problem can be reformulated in terms of the matrix $\Gamma$ (given) and the matrix $M$ (to be found).

PROPOSITION 2 *(see [18]) In the SRM the measurement matrix $M$ is given by the two equivalent expressions*

$$M = T^{-1/2}\,\Gamma\,, \qquad M = \Gamma\, G^{-1/2} \tag{5}$$

*where $T^{-1/2}$ and $G^{-1/2}$ are the inverse square roots (in the Moore–Penrose generalized sense) of $T$ and $G$, respectively.*

The alternative approaches to evaluate $M$, either via $T^{-1/2}$ or via $G^{-1/2}$, are important for efficient computation. We develop these results in the case of GUS, which is of special interest for PPM. The first of (5) gives the reference POVM as $\Pi_0 = T^{-1/2}\,\rho_0\,T^{-1/2}$. Once $T^{-1/2}$ is evaluated, we can obtain the transition probabilities, and then the correct detection probability from (4), which becomes

$$P_c = \mathrm{Tr}\left[(\rho_0 T^{-1/2})^2\right]\,. \tag{6}$$

The Gram matrix approach is less direct. The matrix $G$ is block circulant and its decomposition is related to the discrete Fourier transform (DFT). In [24] we found:

PROPOSITION 3 *In the presence of GUS the $i, j$ block of the Gram matrix, of size $r_0 \times r_0$, can be written in the form*

$$G_{ij} = \gamma_i^* \gamma_j = \gamma_0^* S^{j-i} \gamma_0 = \frac{1}{m} \sum_{k=0}^{m-1} W_m^{k(j-i)} E_k \tag{7}$$

*where $W_m = e^{i2\pi/m}$ and the matrices $E_k$, of order $H$, have the alternative expressions*

$$E_k = m \gamma_0^*\, Y_k\, \gamma_0 = \sum_{i=0}^{m-1} G_{0i} W_m^{-ki}\,. \tag{8}$$

*Here, $Y_k$ are the matrices appearing in the EID of $S$ (see (14)). The correct detection probability is finally given by*

$$P_c = \frac{1}{m} \mathrm{Tr}\Big[\Big\{\sum_{k=0}^{m-1} E_k^{1/2}\Big\}^2\Big]\,. \tag{9}$$

In conclusion, with the Gram matrix approach, the performance evaluation requires finding the square roots $E_k^{1/2}$, by the EID, of the $r_0 \times r_0$ matrices $E_k$ defined by (8).

**SRM optimality with GUS.** Under certain conditions, SRM provides the optimal solution for the general case of mixed states. A (sufficient) condition is that the state constellation has GUS and the reference factors satisfy the relation [21] $\mu_0^*\,\gamma_0 = \alpha\, I_H$, where $\alpha$ is an arbitrary constant. Unfortunately, for PPM this condition holds only for pure states.

## III. QUANTUM PULSE POSITION MODULATION (QPPM)

QPPM is an attractive modulation format, which has recently been considered [22] for possible applications to deep space communications. Indeed, it appears well suited to existing laser modulation techniques, in that it can be regarded as a coded version of the OOK format. In particular the requirements for the transmitting laser are the same as for the OOK coherent modulation, with lower average power. On the other hand, the model of the QPPM we consider below is a paradigm for a whole class of quantum communication scheme (as wavelength modulation). Moreover, it has been proved [3] that the model is a quantum–mechanical equivalent of a general communication system with (classically) orthogonal waveforms.

The performances of QPPM systems are known only in very particular cases. The minimum error probability for QPPM with quantum detection in the absence of thermal noise detection (see below) was found in 1975 by Yuen *et al.*. In 1976 Helstrom [8, Chapter 6] evaluated the error probability in the presence of thermal noise, by assuming that a classical OOK receiver counts the photons slot–by–slot and decides according to a majority rule. In 1983 Dolinar [5] suggested an adaptive measurement approach, called *conditionally nulling*, which achieves very nearly the same error probability of the optimum quantum receiver in the absence of thermal noise. However, no precise evaluation exists in the literature for optimal performances of PPM in the presence of thermal noise.

In QPPM the quantum–mechanical channel is modeled by a composite Hilbert space $\mathcal{H}$, given by the tensor product $\mathcal{H}_0^{\otimes m}$ of $m$ identical Hilbert spaces $\mathcal{H}_0$ of dimension $n$. In this space the transmitter chooses one of $m$ pure quantum states, given by the tensor products

$$|\gamma_i\rangle = |\gamma_{i,m-1}\rangle \otimes \dots \otimes |\gamma_{i1}\rangle \otimes |\gamma_{i0}\rangle \qquad i = 0, \dots, m-1 \tag{10}$$

where $|\gamma_{ij}\rangle = |\gamma^0\rangle$ for $j \neq i$ and $|\gamma_{ij}\rangle = |\gamma^1\rangle$ for $j = i$. In a practical QPPM scheme the two states are assumed to be $|\gamma^0\rangle = |0\rangle$ (the Glauber ground state) and $|\gamma^1\rangle = |\alpha\rangle$ (a Glauber state with an average number of photons $N_s = |\alpha|^2$). The explicit constellation for $m = 4$ was given in (1). If the states are mixed, the composite pure states must be replaced by the composite density operators

$$\rho_i = \rho_{i,m-1} \otimes \dots \otimes \rho_{i1} \otimes \rho_{i0} \qquad i = 0, \dots, m-1 \tag{11}$$

where $\rho_{ij} = \rho^0$ for $j \neq i$ and $\rho_{ij} = \rho^1$ for $j = i$, and $\rho^0$ and $\rho^1$ are assigned density operators on the space $\mathcal{H}_0$.

### A. Symmetry Operator

The GUS structure of the QPPM constellation is apparent. Naively, the state $\rho_{i+1}$ is obtained from $\rho_i$ through a *symmetry*

*operator* $S$ that *pushes each Kronecker factor by one position (modulo* $m$*) to the left*. However, to translate the above words into a formula is not a trivial problem, since the operator $S$ is not separable into a tensor product of operators.

PROPOSITION 4 *The density operators* $\rho_i$ *of QPPM form a GUS constellation, namely* $\rho_i = S^i \rho_0 S^{-i}$*. The symmetry operator is given by*

$$S = \sum_{k=0}^{n-1} w_n(k) \otimes I_H \otimes w_n^*(k) \,, \tag{12}$$

*where* $w_n(k)$ *is the column vector of length* $n$*, whose elements are 0 with exception of a 1 in the* $k$*–th position, and* $I_H$ *is the identity matrix of order* $H = n^{m-1}$.

As an illustrative example, the symmetry operator for $n = 2$ and $m = 4$ (4–PPM) takes the form

$$S = \begin{bmatrix}
1&0&0&0&0&0&0&0&0&0&0&0&0&0&0&0\\
0&0&1&0&0&0&0&0&0&0&0&0&0&0&0&0\\
0&0&0&0&1&0&0&0&0&0&0&0&0&0&0&0\\
0&0&0&0&0&0&1&0&0&0&0&0&0&0&0&0\\
0&0&0&0&0&0&0&0&1&0&0&0&0&0&0&0\\
0&0&0&0&0&0&0&0&0&0&1&0&0&0&0&0\\
0&0&0&0&0&0&0&0&0&0&0&0&1&0&0&0\\
0&0&0&0&0&0&0&0&0&0&0&0&0&0&1&0\\
0&1&0&0&0&0&0&0&0&0&0&0&0&0&0&0\\
0&0&0&1&0&0&0&0&0&0&0&0&0&0&0&0\\
0&0&0&0&0&1&0&0&0&0&0&0&0&0&0&0\\
0&0&0&0&0&0&0&1&0&0&0&0&0&0&0&0\\
0&0&0&0&0&0&0&0&0&1&0&0&0&0&0&0\\
0&0&0&0&0&0&0&0&0&0&0&1&0&0&0&0\\
0&0&0&0&0&0&0&0&0&0&0&0&0&1&0&0\\
0&0&0&0&0&0&0&0&0&0&0&0&0&0&0&1
\end{bmatrix}$$

PROOF. The result is a particular case of a property of the Kronecker product (see [9]). Given two matrices $A$ and $B$, with dimensions $r \times h$ and $s \times k$, respectively, the Kronecker products $A \otimes B$ and $B \otimes A$ are different, although they have the same dimension $rs \times hk$ and contain the same entries, in different positions. Moreover, the products are related by

$$B \otimes A = S_{rs} [\, A \otimes B \,] S_{hk}^{-1} \,, \tag{13}$$

where the matrices $S_{rs}$ and $S_{hk}$ are permutation matrices of order $rs$ and $hk$, respectively, known as *perfect shuffle* matrices, which are independent of the particular matrices $A$ and $B$ and depend only on their dimensions. The matrix $S_{rs}$ is given by $S_{rs} = \sum_{k=0}^{s-1} w_s(k) \otimes I_r \otimes w_s^*(k)$. Now, the symmetry operator of QPPM gives the mapping

$$\rho_{i,m-1} \otimes [\rho_{i,m-2} \otimes \ldots \otimes \rho_{i0}\,] \longrightarrow [\,\rho_{i,m-2} \otimes \ldots \otimes \rho_{i0}\,] \otimes \rho_{i,m-1}$$

which is a particular case of the general transformation (13) with square matrices $A = \rho_{i,m-1}$ of order $n$ and $B = \rho_{i,m-2} \otimes \ldots \otimes \rho_{i0}$ of order $H = n^{m-1}$. By putting $r = h = n$ an $s = k = H$ into (13), we have $S_{rs} = S_{hk} = S_{nH}$ and we find the expression (12).

### B. Eigendecomposition of the Symmetry Operator

As shown above, in the performance evaluation of a quantum detection system that satisfies a GUS constraint, the EID of the symmetry operator $S$ is a key step. Since $S$ is unitary and $S^m = I$, the EID has the form

$$S = \sum_{k=0}^{m-1} \lambda_k \, Y_k \, Y_k^* \tag{14}$$

where the $Y_k$ collect, as their columns, the kets $|y_j\rangle$ corresponding to the eigenvalue $\lambda_k = W_m^{-k}$. The problem is the evaluation of the matrices $Y_k$. We can also exploit the fact that in our case $S$ reduces to a permutation matrix. The EID of permutation matrices is an intricate topic (see for instance [2]), at least for the general case. However, the particular structure (12) of the operator simplifies the results. In any case, the *cycles* generated by $S$, according to the following definition, play a fundamental role for the EID. Provided that $w(k)$, $k = 0, \ldots, n^m - 1$ are the unit vectors of the space $\mathcal{H} = \mathcal{H}^{\otimes n}$, the distinct unit vectors $w(k_0), w(k_1), \ldots, w(k_{p-1})$ form a *cycle* (of *period* $p$) if $S^p w(k_0) = w(k_0)$ and, for each $i = 0, \ldots, p-1$, $w(k_i) = S^i w(k_0)$ .

In Appendix A we discuss the properties of the symmetry operator $S$ listed in the following proposition.

PROPOSITION 5 *1) The periods of the cycles of* $S$ *are divisors of* $m$*. 2) The indexes* $k_0, \ldots, k_{m-1}$ *that identify the cycle* $w(k_0), w(k_1), \ldots, w(k_{p-1})$ *are obtained from* $k_0$ *through the recursive relation*

$$k_{i+1} = n k_i \bmod (n^m - 1) \,. \tag{15}$$

*3) For any cycle* $w(k_0), w(k_1), \ldots, w(k_{p-1})$ *the vector*

$$V_j = \frac{1}{\sqrt{p}} \sum_{h=0}^{p-1} W_p^{jh} w(k_h) \qquad j = 0, \ldots, p-1 \tag{16}$$

*is an eigenvector of* $S$ *associated to the eigenvalue* $W_p^{-j}$*. The eigenvectors so obtained constitute an orthonormal basis of the space* $\mathcal{H}$*.*

As a conclusion, we have the following EID of the symmetry matrix $S$

$$S = \sum_{p\,|\,m} \sum_{c \in \mathcal{K}_p} \sum_{j=0}^{p-1} e^{-i2\pi j/p} V_{cj} V_{cj}^* \tag{17}$$

where the subscript $p\,|\,m$ denotes that the sum is extended to the divisors of $m$ ($m$ included), $\mathcal{K}_p$ is the set of cycles of minimum period $p$ and, for each cycle $c \in \mathcal{K}_p$, the vectors $V_{cj}$, $j = 0, \ldots, p-1$ are the eigenvectors generated by the cycle $c$. After suitably reordering indexes, (17) can be written in the form

$$S = \sum_{i=0}^{m-1} W_m^{-i} Y_i Y_i^* \tag{18}$$

where $Y_i Y_i^*$ turns out to be the projector mapping the space $\mathcal{H}$ into the autospace associated to the eigenvalue $W_m^i$.

In order to evaluate the EID (17), we can use the following proposition, proved in Appendix B.

PROPOSITION 6 *The number* $N_p$ *of unit vectors of minimum period* $p$ *is given by the recursive relation* $n^p = \sum_{d\,|\,p} N_d$*. The number of cycles of period* $p$ *is* $N_p/p$*. The multiplicity of the eigenvalue* $W_m^{-h}$ *is given by* $n_h = \sum_{(m/p)\,|\,h} N_p/p$*,* $h = 1, 2, \ldots, m$*, where the summation is extended to all the* $p$ *such that* $m/p$ *divides* $h$*, and* $n_m$ *gives* $n_0$*.*

### C. Examples

We reconsider the previous example of $S$, for $n = 2$ and $m = 4$ (4–PPM). The possible periods for the cycles are 1, 2, and 4. By applying the recursive relation (15), which becomes $k_{i+1} = 2k_i \mod 15$, we find the cycles $\{0\}$ and $\{15\}$ of period 1, $\{5,10\}$ of period 2, $\{1,2,4,8\}$, $\{3,6,12,9\}$, and $\{7,14,13,11\}$ of period 4.

The eigenvalues $1, W_4, W_4^2, W_4^3$ have 6, 3, 4, and 3 eigenvectors, respectively. The matrices collecting the eigenvectors associated to the eigenvalues $1, W_4, W_4^2, W_4^3$ are then

$$Y_0 = \begin{bmatrix} 1 & 0 & 0 & 0 & 0 & 0 \\ 0 & 0 & 0 & \frac{1}{2} & 0 & 0 \\ 0 & 0 & 0 & \frac{1}{2} & 0 & 0 \\ 0 & 0 & 0 & 0 & \frac{1}{2} & 0 \\ 0 & 0 & 0 & \frac{1}{2} & 0 & 0 \\ 0 & 0 & \frac{1}{\sqrt{2}} & 0 & 0 & 0 \\ 0 & 0 & 0 & 0 & \frac{1}{2} & 0 \\ 0 & 0 & 0 & 0 & 0 & \frac{1}{2} \\ 0 & 0 & 0 & \frac{1}{2} & 0 & 0 \\ 0 & 0 & 0 & 0 & \frac{1}{2} & 0 \\ 0 & 0 & \frac{1}{\sqrt{2}} & 0 & 0 & 0 \\ 0 & 0 & 0 & 0 & 0 & \frac{1}{2} \\ 0 & 0 & 0 & 0 & \frac{1}{2} & 0 \\ 0 & 0 & 0 & 0 & 0 & \frac{1}{2} \\ 0 & 0 & 0 & 0 & 0 & \frac{1}{2} \\ 0 & 1 & 0 & 0 & 0 & 0 \end{bmatrix}, \quad Y_1 = \begin{bmatrix} 0 & 0 & 0 \\ -\frac{i}{2} & 0 & 0 \\ -\frac{1}{2} & 0 & 0 \\ 0 & -\frac{i}{2} & 0 \\ \frac{i}{2} & 0 & 0 \\ 0 & 0 & 0 \\ 0 & -\frac{1}{2} & 0 \\ 0 & 0 & -\frac{i}{2} \\ \frac{1}{2} & 0 & 0 \\ 0 & \frac{1}{2} & 0 \\ 0 & 0 & 0 \\ 0 & 0 & \frac{1}{2} \\ 0 & \frac{i}{2} & 0 \\ 0 & 0 & \frac{i}{2} \\ 0 & 0 & -\frac{1}{2} \\ 0 & 0 & 0 \end{bmatrix}$$

$$Y_2 = \begin{bmatrix} 0 & 0 & 0 & 0 \\ 0 & -\frac{1}{2} & 0 & 0 \\ 0 & \frac{1}{2} & 0 & 0 \\ 0 & 0 & -\frac{1}{2} & 0 \\ 0 & -\frac{1}{2} & 0 & 0 \\ -\frac{1}{\sqrt{2}} & 0 & 0 & 0 \\ 0 & 0 & \frac{1}{2} & 0 \\ 0 & 0 & 0 & -\frac{1}{2} \\ 0 & \frac{1}{2} & 0 & 0 \\ 0 & 0 & \frac{1}{2} & 0 \\ \frac{1}{\sqrt{2}} & 0 & 0 & 0 \\ 0 & 0 & 0 & \frac{1}{2} \\ 0 & 0 & -\frac{1}{2} & 0 \\ 0 & 0 & 0 & -\frac{1}{2} \\ 0 & 0 & 0 & \frac{1}{2} \\ 0 & 0 & 0 & 0 \end{bmatrix}, \quad Y_3 = \begin{bmatrix} 0 & 0 & 0 \\ \frac{i}{2} & 0 & 0 \\ -\frac{1}{2} & 0 & 0 \\ 0 & \frac{i}{2} & 0 \\ -\frac{i}{2} & 0 & 0 \\ 0 & 0 & 0 \\ 0 & -\frac{1}{2} & 0 \\ 0 & 0 & \frac{i}{2} \\ \frac{1}{2} & 0 & 0 \\ 0 & \frac{1}{2} & 0 \\ 0 & 0 & 0 \\ 0 & 0 & \frac{1}{2} \\ 0 & -\frac{i}{2} & 0 \\ 0 & 0 & -\frac{i}{2} \\ 0 & 0 & -\frac{1}{2} \\ 0 & 0 & 0 \end{bmatrix}$$

As another example, let $n = 2$ and $m = 10$, so that $n^m = 1024$. The possible periods for the cycles are $p = 1, 2, 5, 10$. The number of unit vectors with period $p$ are $N_1 = 2$, $N_2 = 2$, $N_5 = 30$, and $N_{10} = 990$. The numbers $N_p/p$ are $2, 1, 6, 99$, respectively. The matrices $Y_i$ have dimension $1024 \times n_i$, where $n_0 = 108$, $n_1 = n_3 = n_7 = n_9 = 99$, $n_2 = n_4 = n_6 = n_8 = 105$, and $n_5 = 100$.

It can be shown that the sizes of the matrices involved increase very rapidly. Fortunately, the matrices $Y_i$ are very sparse and strongly structured. Several appropriate tricks can then be exploited to alleviate the computational complexity.

## IV. Quantum states in coherent optical communications

In quantum optical communication systems the correct setting of quantum states is provided by the Glauber's celebrated theory [1]. In the case of PPM, this theory will provide the basic density operators $\rho^0$ and $\rho^1$, representing the symbols 0 and 1, from which the $m$ density operators $\rho_i$ can be obtained by tensorial products.

### A. Density Operators from Glauber's Theory

In this theory, a *coherent state*, representing a monochromatic electromagnetic radiation produced by a laser, is formulated in an infinite dimensional Hilbert space $\mathcal{H}$ equipped with an orthonormal basis $|n\rangle, n = 0, 1, 2, \ldots$, where $|n\rangle$ are called *number eigenstates*. Each state $|n\rangle$ is said to contain exactly $n$ photons. In this context the Glauber representation of a single radiation mode is given by the ket

$$|\alpha\rangle = e^{-\frac{1}{2}|\alpha|^2} \sum_{n=0}^{\infty} \frac{\alpha^n}{\sqrt{n!}} |n\rangle \tag{19}$$

where $\alpha$ is the complex envelope that specifies the mode and $N_\alpha = |\alpha|^2$ represents the *average number of photons*, when the system is in the coherent state $|\alpha\rangle$.

The representation (19) is valid when the receiver observes a pure state with a known parameter $\alpha$, which in the context of communications may be regarded as the *signal*. In the presence of thermal (or background) noise, the signal becomes uncertain and must be represented through a density operator. In this case the Glauber theory provides an infinite matrix representation $||\rho_{mn}||$ of the density operator, given (for $0 \le m \le n$) by

$$\begin{aligned} \rho_{mn}(\alpha) = (1-v)v^n \sqrt{\frac{m!}{n!}} \left(\frac{\alpha^*}{\mathcal{N}}\right)^{n-m} e^{-(1-v)|\alpha|^2} \cdot \\ \cdot L_m^{n-m}\left(-\frac{|\alpha|^2}{\mathcal{N}(\mathcal{N}+1)}\right) \end{aligned} \tag{20}$$

where $\mathcal{N}$ represents the *average number of photons* associated with the thermal noise, $v = \mathcal{N}/(1+\mathcal{N})$, and $L_m^{n-m}(x)$ are generalized Laguerre polynomials. The entries for $m > n$ are obtained by $\rho_{nm}(\alpha) = \rho_{mn}^*(\alpha)$. The diagonal elements $\rho_{mm}(\alpha)$ obey a Laguerre distribution and give the probabilities of exactly $m$ photons being present, when the quantum system is in the noisy state represented by the density operator $\rho(\alpha)$ [3]. For the ground state $|\gamma\rangle = |0\rangle$ the above expression degenerates and the corresponding matrix representation becomes diagonal, namely $\rho_{mn}(0) = \delta_{mn}(1-v)v^n$.

The infinite dimensional matrix $||\rho_{mn}(\alpha)||, 0 \le m, n < \infty$ gives a correct representation of the density operator, but, for practical calculations, we introduce a finite dimensional approximation by truncating to $n_\epsilon$ terms. For the choice of $n_\epsilon$, to achieve a given accuracy, we follow the *quasi–unitary trace criterion*, proposed in [24], which is based on the fact that a density operator has unitary trace. Then, we choose $n_\epsilon$ as the smallest integer, such that $\sum_{m=0}^{n_\epsilon - 1} \rho_{mm}(\alpha) \ge 1 - \epsilon$, where $\epsilon$ is the required accuracy. Thus, for a given $\epsilon$, $n_\epsilon$ can be evaluated using the Laguerre distribution $\rho_{mm}(\alpha)$.

**Numerical accuracy.** The above numerical approach implies a twofold approximation. First, the truncation of the infinite matrix to a finite $n_\epsilon \times n_\epsilon$ matrix $\rho$. It is reasonable to think that, at the increasing of $n_\epsilon$ , the approximate results converges to the correct ones. However, the evaluation of the corresponding error seems to be a difficult problem and we made no attempt to evaluate it. A second approximation is that the truncated density matrix is not normalized to unitary trace. Of course, it can be normalized as $\bar{\rho} = c\,\rho$, where $c = 1/\mathrm{Tr}(\rho) \le 1/(1-\epsilon)$. In the PPM the truncation is operated on the basic operators $\rho^0$ and $\rho^1$, which we normalize as $\bar{\rho}^0 = c\,\rho^0$ and $\bar{\rho}^1 = c\,\rho^1$. To see the non normalization effect on the error probability $P_e = 1 - P_c$, we consider the SRM approach stated by Proposition3, and mark with a bar the quantities obtained from normalized density operators. For the reference factor we get $\bar{\gamma}_0 = c^{m/2}\,\gamma_0$ and for the square–root matrices $\bar{E}_k^{1/2} = c^{m/2}\,E_k^{1/2}$, and finally from (9) $\bar{P}_c = c^m\,P_c$. Hence, $\bar{P}_e = (1 - c^m) + c^m\,P_c$ and, considering that $\epsilon \ll 1$ and $P_e \ll 1$, we find $P_e - \bar{P}_e \simeq m\epsilon$. In conclusion, if we choose $\epsilon \ll P_e$, we see that normalization is irrelevant. In the numerical evaluation of Section V, where $P_e > 10^{-6}$, we have chosen $\epsilon = 10^{-8}$.

### *B. Factorization of the density operators. Choice of the rank*

Once the finite $n_\epsilon \times n_\epsilon$ approximation of the density operator has been established, we need a factorization of the form $\rho(\alpha) = \gamma(\alpha)\gamma^*(\alpha)$ for a suitable matrix $\gamma(\alpha)$. For $\alpha = 0$ (ground state) the factorization is immediate, since $\rho(0)$ is diagonal, and we find $\gamma(0) = \sqrt{\rho(0)} = ||\delta_{mn}\sqrt{(1-v)v^n}||$. For $\alpha \neq 0$ we evaluate the reduced EID of $\rho(\alpha)$, namely $\rho(\alpha) = Z_h \Lambda_h^2 Z_h^*$, where $h$ is the rank of $\rho(\alpha)$, $Z_h$ is an $n_\epsilon \times h$ matrix that collects the eigenvectors corresponding to the $h$ positive eigenvalues $\lambda_i^2$, and $\Lambda_h^2$ is the $h \times h$ diagonal matrix that collects the $\lambda_i^2$. Hence, $\gamma(\alpha) = Z_h \Lambda_h$ is a factor of $\rho(\alpha)$.

A critical step in the numerical evaluation is the choice of the rank $h$, given by the number of numerically relevant eigenvalues. To clarify the problem we develop a specific case: $\alpha = \sqrt{5}$ , $N_s = 5$ , $\mathcal{N} = 0.1$ , $\epsilon = 10^{-5} \quad \rightarrow \quad n_\epsilon = 20$. Now, in theory, $\rho(\alpha)$ has full rank $h = n_\epsilon$, as can be seen from the list of its eigenvalues, obtained with a high accuracy

$$0.150285\,,\quad 0.00231095\,,\quad 0.0000353779\,,\quad 5.20725\,10^{-7}$$
$$7.24874\,10^{-9}\,,\quad 9.45157\,10^{-11}\,,\quad 1.14603\,10^{-12}\,,\quad \ldots$$
$$3.10867\,10^{-21}\,,\quad 2.32186\,10^{-21}\,,\quad 1.37631\,10^{-22}\,,\quad 3.01775\,10^{-25}$$

but in practice, we can limit the list to the first 3 eigenvalues, neglecting the rest, which means that we assume $h = 3$ as a "practical" rank. As a check, the reconstruction of $\rho(\alpha)$ obtained in this way assures an accuracy $< 10^{-8}$. To find the "practical" rank in the general case, we consider the reconstruction error $\Delta\rho = \rho - \gamma_h\gamma_h^*$, where the factor $\gamma_h$ is obtained by taking only $h$ eigenvalues. Then, we can evaluate the maximum error, or the mean square error, as a function of $h$ and choose $h = h_\nu$ to achieve a given accuracy $\nu$ (for a detailed discussion of $n_\epsilon$ and $h_\nu$ versus the accuracies $\epsilon$ and $\nu$, see [24]).

## V. Application examples

We apply the theory of the previous sections to evaluate the error probability in QPPM systems. But first we compare the developed methods to give a guide to possible ranges in the different numerical evaluations.

### *A. Comparison of computational methods*

In the previous section, we saw that, given the accuracies $\epsilon$ and $\nu$, we can evaluate the basic density operators $\rho^0 = \rho(0) = \gamma_h(0)\gamma_h(0)^*$ and $\rho^1 = \rho(\alpha) = \gamma_h(\alpha)\gamma_h(\alpha)^*$ for any choice of signal parameter $N_s = |\alpha|^2$ and of thermal noise parameter $\mathcal{N}$ (it is expedient to choose the same "practical" rank $h$ both for $\rho^0$ and $\rho^1$). Hence, the dimensions and the ranks are as follows:

- $\gamma^0$ and $\gamma^1$ $\quad n \times h \quad$ rank $h$, $\quad \rho^0$ and $\rho^1$ $\quad n \times n \quad$ rank $h$,
- $\gamma_i$ $\quad N \times H \quad$ rank $H$, $\quad \rho_i$ $\quad N \times N \quad$ rank $H$, with $N = n^m$ and $H = h^m$,
- state matrix $\Gamma$ $\quad N \times mH \quad$ rank $r = mH$ (if $mH \leq N$),
- Gram operator $T$ $\quad N \times N \quad$ rank $mH$, $\quad$ Gram matrix $G$ $\quad mH \times mH \quad$ rank $mH$.

Having this schedule in mind, we now compare the different evaluation methods, considering that the practical limit is determined by the dimensions of the input data in convex linear programming (CLP), and that in general, it is given by the dimensions of the matrices to be eigen–decomposed.

*1) Optimal solution:* This solution is available for:

- pure states ($\mathcal{N} = 0$), for any order $m$,
- mixed states, for $m = 2$ from Helstrom's theory, but an $N \times N$ EID is required,
- CLP with input the $N \times N$ density operators $\rho_i$.

*2) SRM solution:* This solution is always available, for any order $m$, for both pure and mixed states. We have two possible approaches

- via Gram operator, requires the EID of an $N \times N$ matrix,
- via Gram matrix, requires the EIDs of $H \times H$ matrices.

Since $H \ll N$, we have used the Gram matrix approach, with a practical limit for the EID computation of $H \simeq 1500$. With the CLP (Matlab toolbox LMI), we found the practical limit of the size $N$ of the input density operators to be no greater than 150. These limits refer to a standard personal computer.

To complete the preliminaries we consider two more topics.

**Error probability with pure states.** We saw that SRM gives the optimal solution with pure states, by virtue of the PPM's GUS. For the explicit evaluation, we use (8), where $G_{0s} = \gamma_0^*\gamma_s = \langle\gamma_0|\gamma_s\rangle = \langle\gamma_{00}|\gamma_{s0}\rangle \cdots \langle\gamma_{0,m-1}|\gamma_{s,m-1}\rangle$. Recalling that the inner product of two Glauber states is given by [1] as $\langle\alpha|\beta\rangle = \exp(-\frac{1}{2}(|\alpha|^2 + |\beta|^2 - 2\alpha^*\beta))$, we get $G_{0s} = 1$ for $s = 0$ and $G_{0s} = \exp(-|\alpha|^2) = \exp(-N_s)$ for $s \neq 0$. Hence, from (9), we have

$$P_c = \frac{1}{m^2}\left(\sqrt{1 + (m-1)\,e^{-N_s}} + (m-1)\sqrt{1 - e^{-N_s}}\right)^2 \tag{21}$$

which is in perfect agreement with the result of [7], obtained with a different approach.

**Comparison with classical optical PPM.** In a classical optical PPM system based on a photon counter, the error probability was evaluated by Helstrom [8, Chapter VI], using the Laguerre distribution of photon arrivals. He found

$$P_e = \frac{1}{m}\sum_{i=2}^{m}(-1)^i\binom{m}{i}\exp\left[-\frac{(1-v)(1-v^{i-1})\,N_s}{1-v^i}\right]$$

where $v = \mathcal{N}/(1+\mathcal{N})$.

### *B. Application to 2–PPM*

For $m = 2$, we do not encounter numerical problems because, for the range of interest, the dimension are relatively small. We first present the results in the absence of thermal noise (pure states), compared with the performance of a classic optical PPM system (Fig. 1), which confirms the substantial improvement of the quantum system with respect to the classical one.

In the presence of thermal noise, for 2–PPM (as for any other binary format) an exact evaluation is possible using Helstrom's theory (see Proposition 1). This possibility was also used to check the results obtained with the Matlab LMI toolbox. For SRM we used the Gram matrix approach. The results are shown in Fig. 2 for some values of the thermal noise parameter $\mathcal{N}$ and show that SRM overestimates the error probability by about 30%, with respect to the optimum. We conclude that SRM is also "pretty good" in the presence of thermal noise.

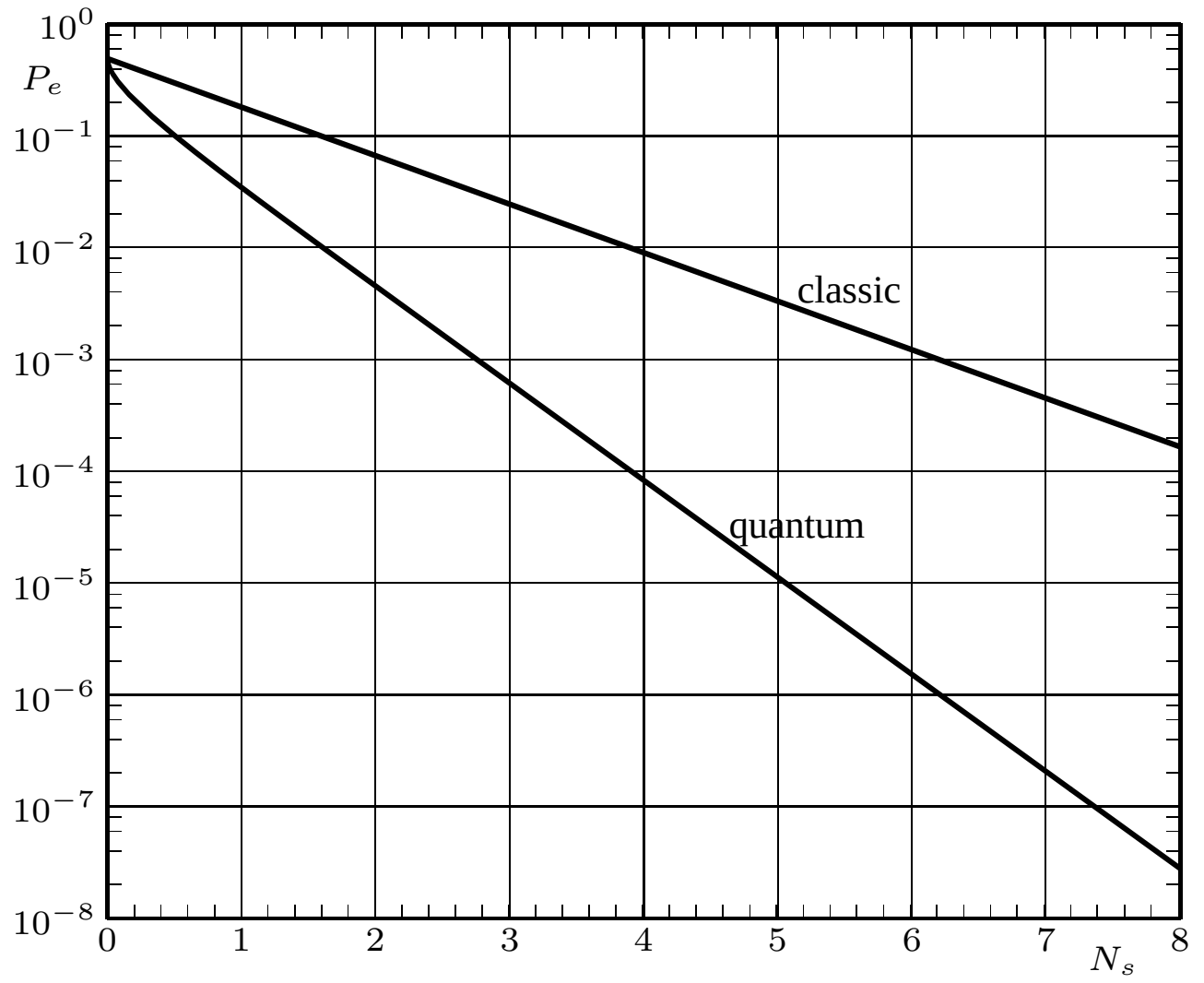

Fig. 1 — Error probability of classic and quantum 2–PPM in the absence of thermal noise. $N_s$ is the average number of photons per symbol.

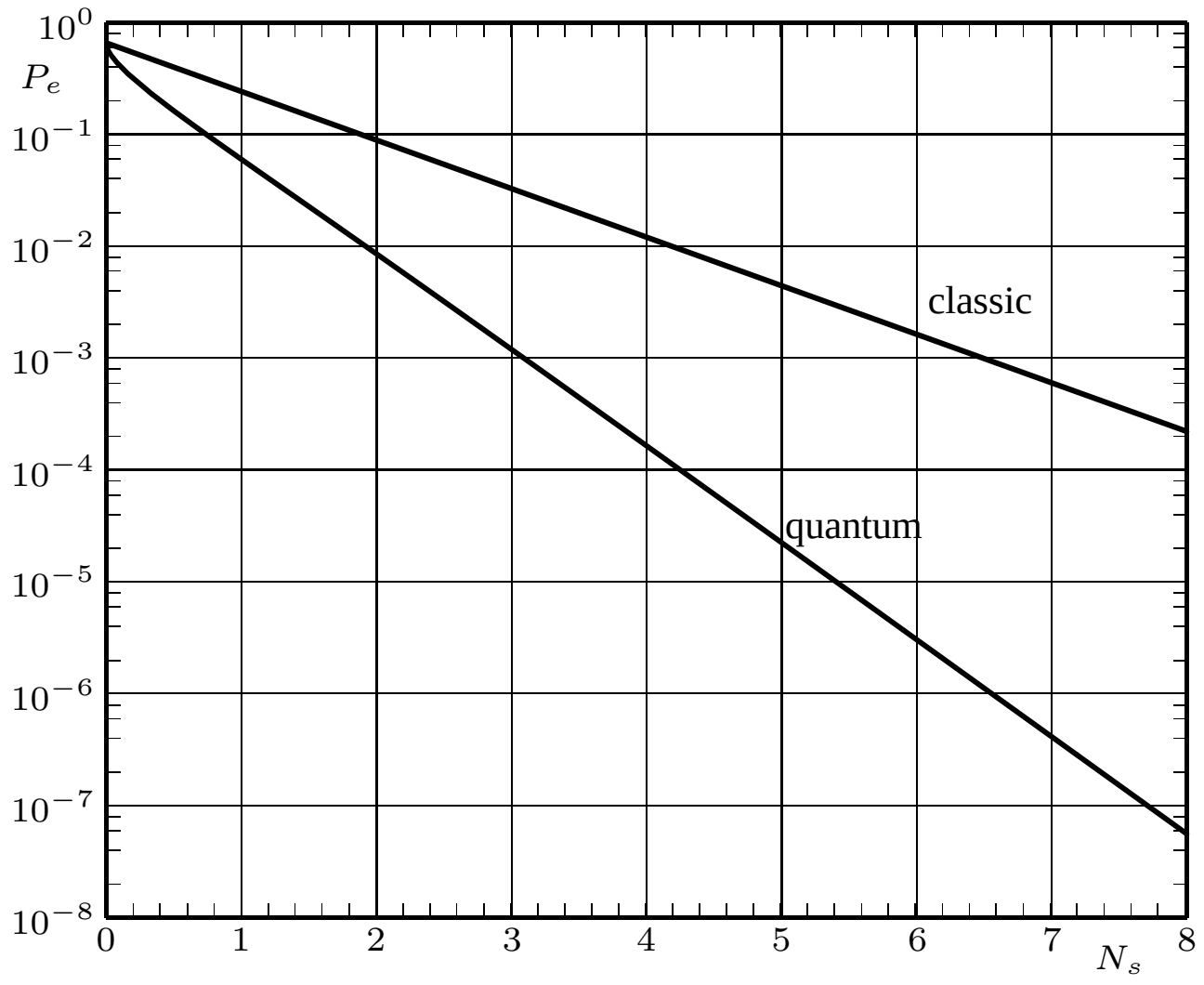

Fig. 3 — Error probability versus the average number of photons per symbol $N_s$ of classic and quantum 3–PPM in the absence of thermal noise.

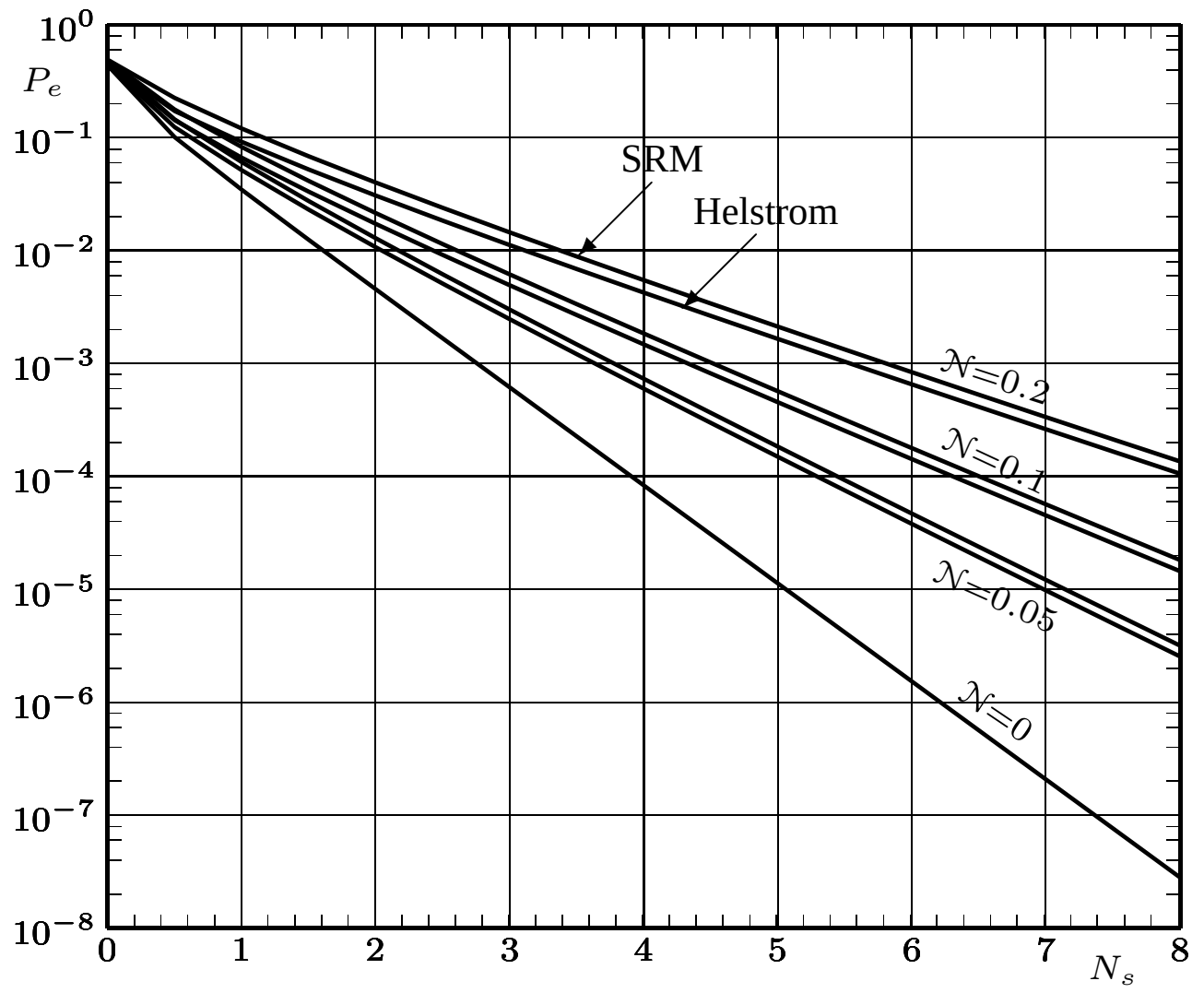

Fig. 2 — Error probability in 2–PPM versus the average number of photons per symbol $N_s$ for some values of the thermal noise parameter $\mathcal{N}$. Comparison of SRM results with Helstrom's optimum results.

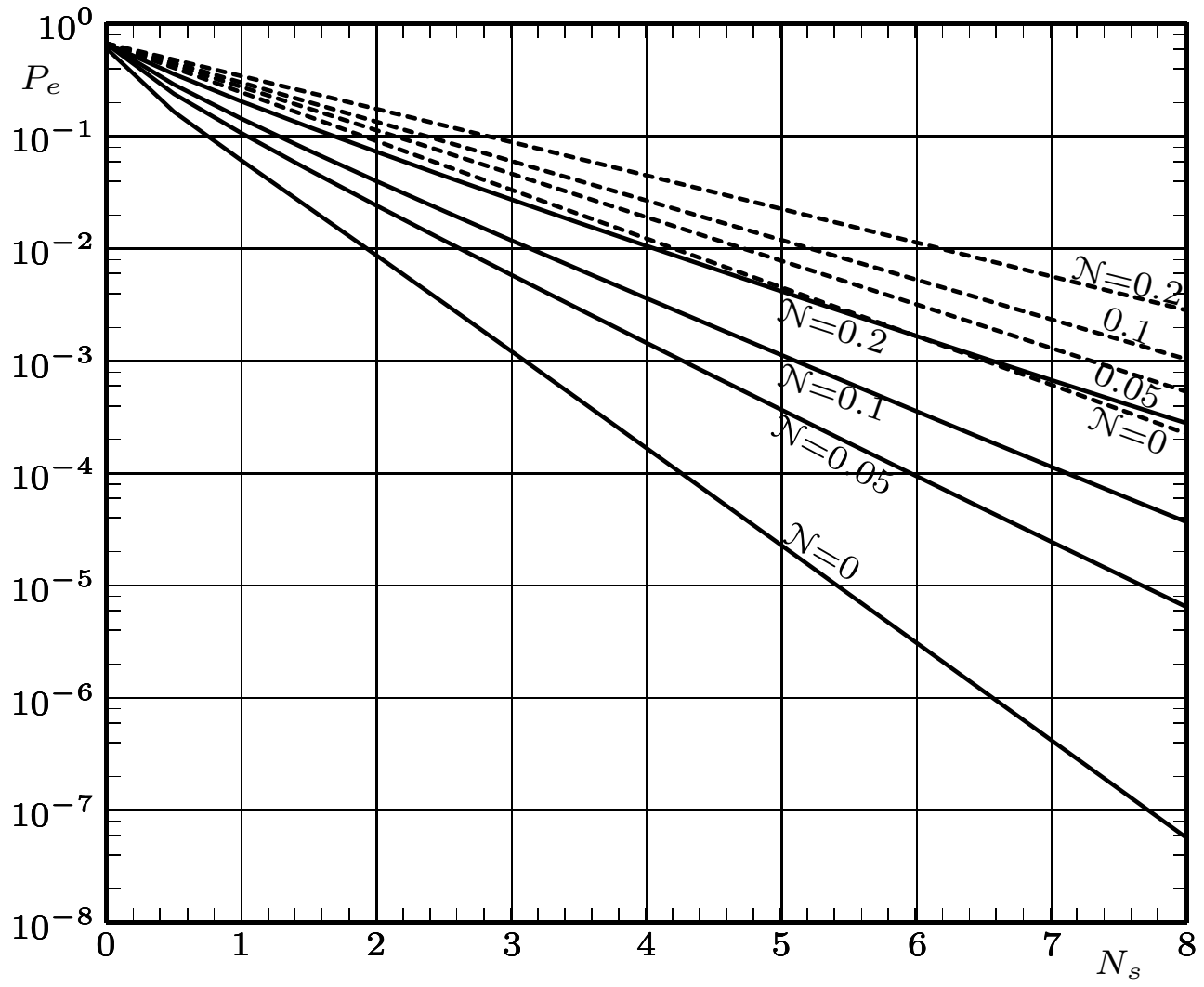

Fig. 4 — Error probability in 3–PPM versus the average number of photons per symbol $N_s$ for some values of the thermal noise parameter $\mathcal{N}$. Solid lines refer to quantum detection and dashed lines to classical photon counter detection.

### *C. Application to 3–PPM*

For $m = 3$, we encounter some numerical problems that can be overcome using the alternative approaches. We first present our results in the absence of thermal noise (pure states), compared with the performance of a classic optical PPM system (Fig.3). We realize the considerable improvement of the quantum system with respect to the classical one.

In the presence of noise, we used the SRM Gram matrix approach. The results are shown in Fig. 4 as a function of $N_s$, for some values of the thermal noise parameter $\mathcal{N}$. To extend the error probability range down to $10^{-5}$ we chose $n = 40$, $h = 8$, hence $H = 8^3 = 512$, which is within the limit of our computational capability. The figure also compares quantum detection with classical photon counter detection, and shows more than 3 decades improvement in error probability.

Since SRM is not optimal in the presence of noise, for the 3–PPM we tried to evaluate the minimum error probability by CLP. As shown in Fig. 5, this was feasible for a very small range of $N_s$, where the dimension $N = n^3$ can be kept below a value of 150. Anyway, this has permitted to show that the SRM gives a very small overestimate of the error probability.

### *D. Application to 4–PPM*

For $m = 4$, the numerical problems become very severe and the comparison with the optimal case (via CLP) is not

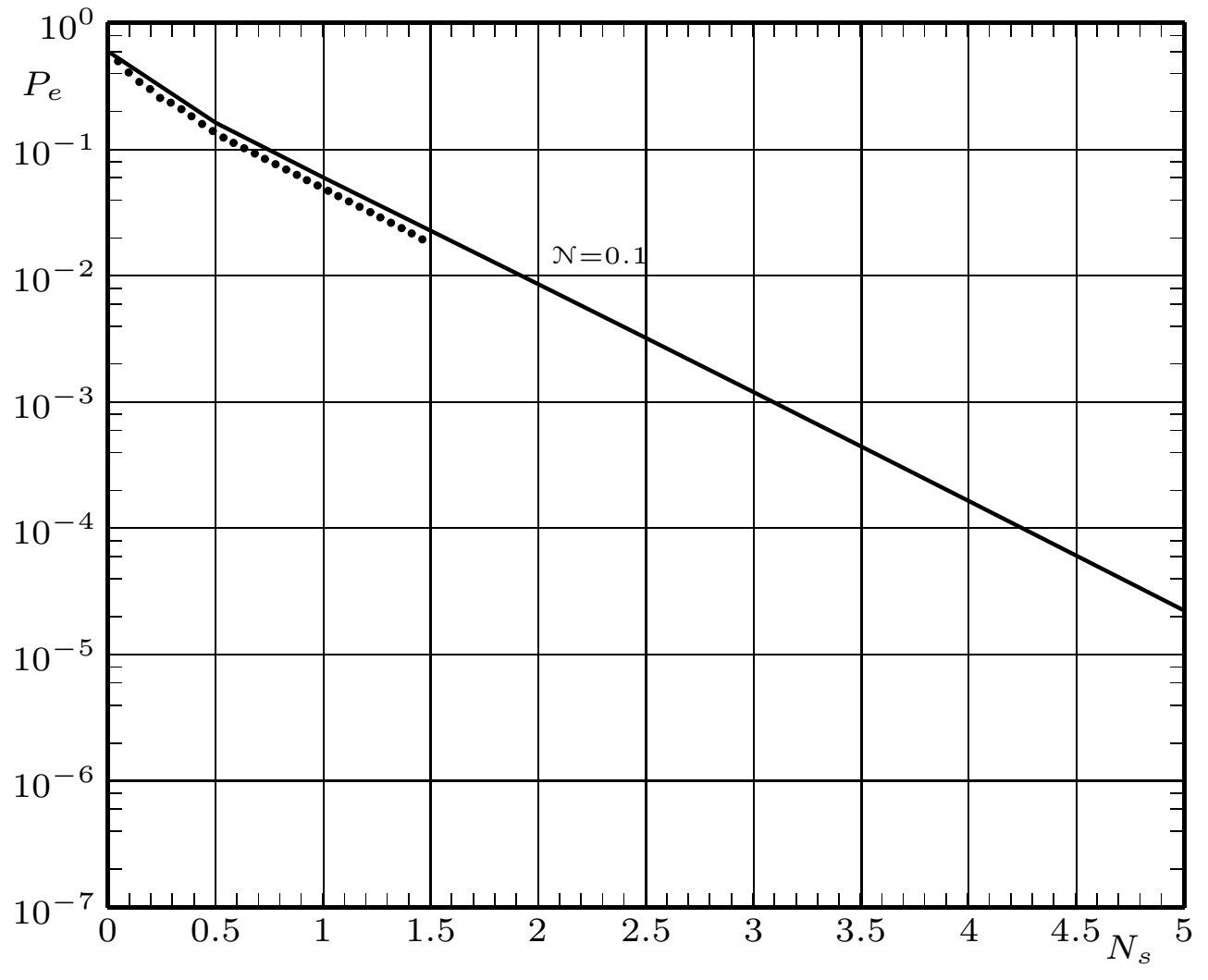

Fig. 5 — Error probability in 3–PPM versus the average number of photons per symbol $N_s$ for the value $\mathcal{N} = 0.1$ of the thermal noise parameter. Solid line refers to SRM computation and dotted lines to convex linear programming (CLP) computation.

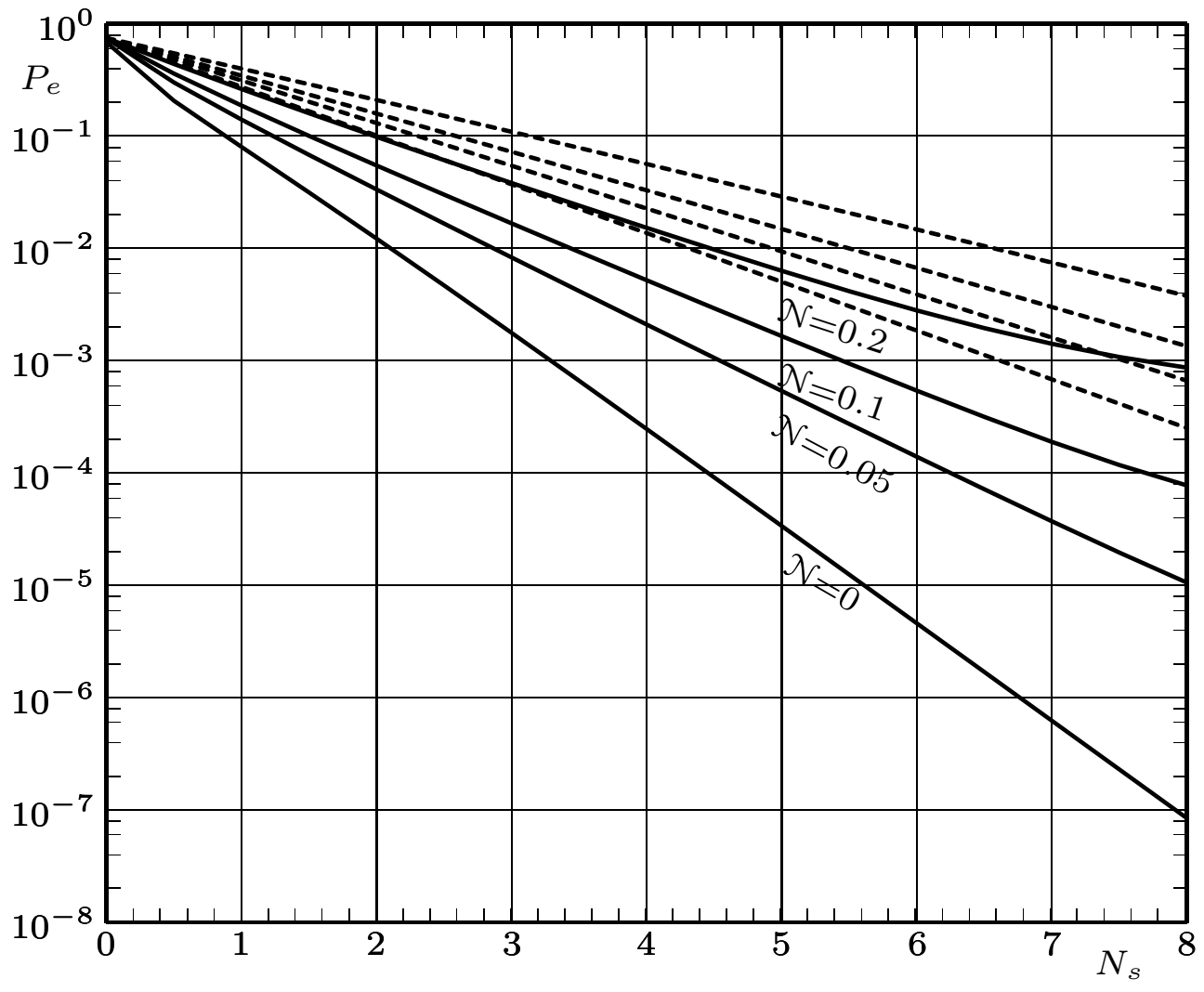

Fig. 7 — Error probability in 4–PPM versus the average number of photons per symbol $N_s$ for some values of the thermal noise parameter $\mathcal{N}$. Solid lines refer to quantum detection and dashed lines to classical photon counter detection.

possible. We present the results in the absence of thermal noise (pure states), compared with the performance of a classic optical PPM system in Fig. 6 and we realize the considerable

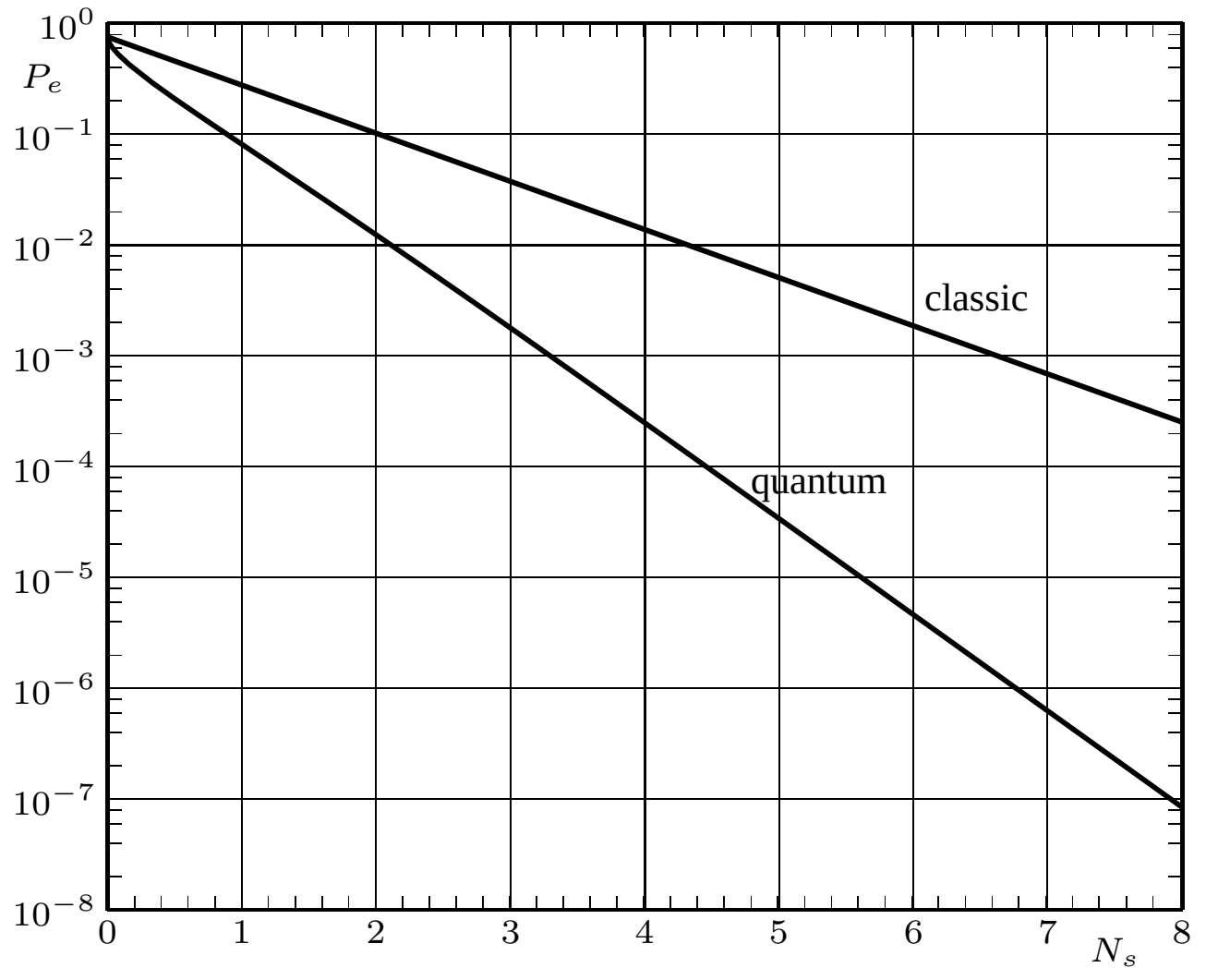

Fig. 6 — Error probability versus the average number of photons per symbol $N_s$ of classic and quantum 4–PPM in the absence of thermal noise.

improvement of quantum 4–PPM with respect to the classical one.

In the presence of noise, we used the SRM Gram matrix approach. The results are shown in Fig. 7 as a function of $N_s$ for some values of the thermal noise parameter $\mathcal{N}$. To cover the error probability range down to $10^{-5}$ we chose $n = 30$, $h = 6$, hence $H = 6^4 = 1296$, which is at the limit of our computation capability. The figure also compares quantum detection with classical photon counter detection and shows more than 3 decades improvement in error probability.

## VI. On the Implementation

The problem of designing and implementing an optimal quantum receiver is a very difficult one also in the simplest cases, as witnessed by the history of the optimal detector for the quantum OOK modulation format. As known, the error probability that can be obtained with a classical detector (direct photon counter) in the absence of thermal noise is given $P_e = (1/2)e^{-2N_s}$, while the error probability achievable by optimal quantum detection (the Helstrom bound) is $P_{\text{opt}} = (1/2)\left[1 - \sqrt{1 - e^{-2N_s}}\right]$ with asymptotical approximation $P_{\text{opt}} \approx (1/4)e^{-2N_s}$.

In 1973 Dolinar [5] proposed an optimum receiver ideally achieving the Helstrom bound. Under the Dolinar's receiver it was the idea of using an adaptive detection, by adding to the received coherent field a time–varying feedback controlled field. Owing to the lack of an adequate technology in fast electro–optics, the realization of the Dolinar receiver was not possible, until in 2007 Cook *et al.* [23] demonstrated its feasibility in a significant range of physical parameters.

For a quantum PPM a natural solution could be to use an OOK receiver for a slot–by–slot hard detection slot by slot. For pure quantum states $|0\rangle$ and $|\alpha\rangle$, a classical photon counter receiver may fail only in detecting the state $|\alpha\rangle$. It is easy to show that the error probability turns out to be $P_e = (m/(m-1))e^{-N_s}$. This is an exponentially degraded performance with respect to the quantum limit (21) having asymptotical approximation $P_{\text{opt}} \approx (1/4)(m-1)e^{-2N_s}$.

On the other hand, an algorithm devised by Dolinar and called "conditionally nulling" [10] allows to obtain an asymptotical error probability $P_{\text{opt}} \approx (1/2)(m-1)e^{-2N_s}$, only double of the quantum limit. This near–optimum algorithm decides at each slot (on the basis of the previous results) whether or not to add to the incoming optical signal a negative replica of the signal corresponding to the quantum state $|\alpha\rangle$.

Unfortunately, the above considerations hold true only in the absence of thermal noise. The degradation due to thermal noise is object of research in progress.

## VII. Conclusion

We have examined the quantum PPM format, both in the absence (pure states) and in the presence (mixed states) of thermal noise, using the Glauber representation of coherent states. Our target was a numerical evaluation of the performance of 4–PPM, where the dimension of the Hilbert space reaches ten thousand and more. To this end, we explored all the currently available methods, namely: exact solution, convex linear programming (CLP), and square root measurement (SRM), and also exploited the geometrically uniform symmetry (GUS), a property not considered before for the PPM, with the goal to find the least expensive implementable solution. We succeeded in this, but just at the limit, so that the methods used do not appear to be useful for higher orders, e.g., 8–PPM. The use of CLP was limited to a small error probability range in 3–PPM.

Whenever possible, we compared SRM with optimal detection and concluded that SRM is very close to optimality. Comparisons with the performance of classical detection confirms the superiority of the quantum detection also when the thermal noise is present.

### *Appendix A: Proof of Proposition 5*

Given a unit vector $w(k)$ of the composite Hilbert space $\mathcal{H} = \mathcal{H}_0^{\otimes n}$, a positive integer $q$ is a *period* of $w(k)$ (with reference to the operator $S$) if $S^q w(k) = w(k)$. The minimum $p$ of periods of $w(k)$ is its *minimum period*. Simple considerations guarantee some obvious facts, such as the existence of a minimum period for each unit vector $w(k)$ and the fact that $S^q w(k) = w(k)$, if and only if, $p$ is a divisor of $q$. In particular, since $S^m w(k) = w(k)$, the minimum period $p$ of each unit vector must be a divisor of $m$.

If we use the $n$–ary representation of the integer $k$, namely, $k = k_{m-1}n^{m-1} + \ldots + k_1 n + k_0$, with $0 \le k_i < m$, the permutation

$$\begin{aligned} S[\, w_n(k_{m-1}) \otimes \ldots \otimes w_n(k_1) \otimes w_n(k_0)\,] \\ = w_n(k_{m-2}) \otimes \ldots \otimes w_n(k_0) \otimes w_n(k_{m-1}) \end{aligned} \quad (22)$$

can be written in the form $Sw(k) = w(h)$, where $h = k_{m-2}n^{m-1} + \ldots + k_0 n + k_{m-1}$. Since

$$nk = \sum_{i=0}^{m-2} k_i n^{i+1} + k_{m-1} + k_{m-1}(n^m - 1) = h \bmod (n^m - 1) \,,$$

the recursion (15) holds.

Since the vectors (16) satisfy the condition

$$\begin{aligned} SV_j &= \frac{1}{\sqrt{p}} \sum_{h=0}^{p-1} e^{i2\pi jh/p} Sw(k_h) \\ &= \frac{1}{\sqrt{p}} \left[ \sum_{h=0}^{p-2} e^{i2\pi jh/p} w(k_{h+1}) + e^{i2\pi j(p-1)/p} w(k_0) \right] \\ &= e^{-i2\pi j/p} V_j \end{aligned}$$

$V_j$ is an eigenvector of $S$ associated to the eigenvalue $W_p^{-j} = e^{-i2\pi j/p}$. Since $w_{k_i}^* w_{k_j} = \delta_{ij}$, it can be immediately verified that the eigenvectors corresponding to a single cycle are orthonormal. Moreover, eigenvectors corresponding to different cycles are orthogonal, in that they are linear combinations of disjoint sets of unit vectors. Then, item 3) follows.

### *Appendix B: Proof of Proposition 6*

By examining the permutation (22), it turns out that the unit vector $w(k)$ has period $d$ (not necessarily minimum), if and only if, the $n$–ary $k_0 k_1 \ldots k_{m-1}$ representation of the index $k$ is periodic, of period $d$, with respect to the cyclic shift, i.e., it is given by the concatenation of $m/d$ equal subsequences $k_0 k_1 \ldots k_{d-1}$. Since the number of such subsequences is $n^d$, this is also the number of unit vectors with period $d$. Since a unit vector has period $d$, if and only if, its minimum period divides $d$, the recursion $n^p = \sum_{d\,|\,p} N_d$ follows. Moreover, for each subsequence $k_0 k_1 \ldots k_{d-1}$, its $d$ cyclic shift give rise to the same cycle, so that the number of cycles of period $d$ is $N_d/d$. Finally, the multiplicity is stated from the fact that a cycle of period $p$ gives rise to an eigenvector associated to $W_m^{-h}$, if and only if, there exists an integer $r$ such that $W_m^{-rm/p} = W_m^{-h}$, i.e., if and only if, $h$ is a multiple of $m/p$.

### *Acknowledgment*

The authors wish to thank A. Assalini, R. Corvaja, and P. Kraniauskas for stimulating discussions and helpful comments.